\documentclass[11pt]{article}
\usepackage[fleqn]{amsmath}

\begin{document}
\setcounter{equation}{0}
\setcounter{figure}{0}
\def\mib#1{\mbox{\boldmath $#1$}}

\begin{flushright}
arXiv:0707.4554 [hep-th]

August 7, 2007

OU-HET 583

\end{flushright}

\vskip1cm
\begin{center}
{\Large{\bf 
Exact  Solutions and the Attractor Mechanism

in Non-BPS Black Holes 
}}
\end{center}

\vskip2.5cm
\begin{center}
{\large {\bf Kyosuke Hotta
\footnote{hotta@het.phys.sci.osaka-u.ac.jp}
 and Takahiro Kubota
 \footnote{kubota@het.phys.sci.osaka-u.ac.jp}
  }}

\vskip0.2cm
{\it Graduate School of Science, Osaka University,

Toyonaka, Osaka 560-0043, Japan}
\end{center}

\vskip1cm
\begin{abstract}
The attractor mechanism for the four-dimensional 
${\cal N}=2$ supergravity black hole 
solution is analyzed in the case of the D0-D4 system. 
Our analyses are based  on   newly derived  exact solutions,  
which exhibit explicitly the attractor mechanism for extremal 
non-BPS black holes. 
Our solutions account for the moduli 
as general complex fields, while in almost all non-BPS solutions obtained previously, the moduli fields are restricted to be purely imaginary. 
It is also pointed out  that our moduli 
solutions contain an extra 
parameter that is not contained in solutions  obtained by replacing the charges  in the  
double extremal  moduli solutions   by the corresponding harmonic functions. 
\end{abstract}
%
\vfill\eject
\section{Introduction}\label{Intro}
\setcounter{equation}{0}
\setcounter{figure}{0}
\renewcommand{\theequation}{1.\arabic{equation}}
\renewcommand{\thefigure}{1.\arabic{figure}}

It has been pointed out that  supersymmetric (SUSY) black hole 
solutions exhibit a peculiar property called the attractor mechanism  
\cite{fks}-\cite{denef}.
It has been confirmed in the case of extremal black holes  that 
moduli fields  are drawn to some fixed values at the horizon 
of the black holes, independently of their asymptotic values. 
In other words, the fixed values of 
the moduli at the horizon are characterized only by the charges carried by the 
black holes. This fact has been studied  by using the BPS  attractor 
equations, 
and an algorithm for calculating the macroscopic Bekenstein-Hawking entropy 
has been established. As a result, it has been found that the entropy is given by the extremum value of the 
central charge \cite{fk}.

In the last several years, the study of the 
attractor mechanism has been extended to non-supersymmetric cases 
\cite{gijt}-\cite{saraikin}, \cite{ccdo}.
Many of the properties of attractive BPS configurations seem to be shared 
by non-BPS attractor configurations, provided that the solutions are extremal. 
A non-BPS attractor equation has  been constructed to relate the charges 
to the attractive values of the moduli \cite{nonbpsattractor}. 
Although the attractor equation is very 
useful, the most direct way 
to examine the nature of black hole solutions is to obtain the solution for 
the moduli fields in the whole space. 
However, the analytic approach to obtaining solutions in non-BPS cases 
is complicated, because in such cases, it is necessary to deal with second-order differential equations, while the 
BPS equations in SUSY cases are first order.
In \cite{gijt}, a  perturbative method is applied to 
extremal black holes, and it is found that the attractor is effective. 
Numerical results support these perturbative results.

It is pointed out in  \cite{sabra}, \cite{denef} and \cite{mohaupt} 
 that in the BPS case, the exact supersymmetric  solutions 
of moduli fields in ${\cal N}=2$ supergravity coupled 
to an arbitrary number of vector multiplets 
may be obtained from the double extremal solutions by simply  replacing 
the charges by harmonic functions. In the case of non-BPS black holes with 
D2 and D6 brane charges,  Kallosh et al. \cite{kallosh3} remarked that the 
exact solution in the STU model can be obtained using the same procedure of 
replacing the charges in the double extremal moduli solutions  
by harmonic functions. 
Their solutions are, however, still restricted, because  the moduli 
fields are 
not general complex numbers but, rather, purely imaginary ones. 

The purpose of the present paper is to derive general exact solutions 
in four-dimensional ${\cal N}=2$ supergravity 
(from the Type IIA superstring)
coupled to vector multiplets for the case 
of a non-BPS extremal black hole with D0-D4 brane charges. 
Our exact solutions obtained for the STU model are more 
general than those obtained by previous authors
for the  D0-D4 system, 
because the moduli fields 
are not restricted to purely imaginary values,  but, instead, are general complex numbers. 
As it turns out, the initial conditions of our solutions have more degrees of freedom  than  those obtained by simply replacing the charges 
in the double extremal moduli solutions 
by the corresponding  harmonic functions. 
The additional degrees of freedom come from the arbitrary complex values 
that we choose for the asymptotic values of the real parts of the  moduli fields.

In a recent paper \cite{saraikin}, 
Saraikin and Vafa  obtained non-BPS double extremal 
black hole solutions, treating the scalar fields as general 
complex numbers.  It is a very intriguing question whether the 
prescription of replacing 
the charges 
at the horizon in their solution by the corresponding harmonic 
functions leads  to general solutions in the whole space. For the BPS case, it has been shown that this prescription is in fact effective for obtaining general 
solutions (see Appendix \ref{bps} for more details). It is beyond the 
scope of the present paper to give an answer to the above question, but 
the explicit non-BPS extremal solutions derived in this paper could be 
useful for investigating such questions.

\section{${\cal N}=2$ supergravity}\label{}
\setcounter{equation}{0}
\setcounter{figure}{0}
\renewcommand{\theequation}{2.\arabic{equation}}
\renewcommand{\thefigure}{2.\arabic{figure}}

We study ${\cal N}=2$ supergravity coupled to $(N_{V}+1)$ 
vector multiplets. 
The bosonic part of the Lagrangian is given by 
\begin{eqnarray}
8\pi e^{-1}{\cal L}=
-\frac{1}{2}R-G_{a\bar b}\partial _{\mu}z^{a}\partial ^{\mu}\bar z^{b}
+\frac{i}{4}\left (
{\overline {\cal  N}}_{IJ}F^{I-}_{\mu \nu}F^{J-\mu \nu}
-{\cal N}_{IJ}F^{I+}_{\mu \nu}F^{J+\mu \nu}
\right ).
\label{eq:lagrangian}
\end{eqnarray}
Here, we define the moduli fields 
\begin{eqnarray}
z^{a}=\frac{X^{a}}{X^{0}}, \hskip0.5cm (a=1,2, \cdots , N_{V}) \hskip0.5cm 
z^{0}=1
\end{eqnarray}
in terms  of the complex scalar field $X^{I} \: 
(\textrm{where}\,\,\, I=0, 1, \cdots , N_{V})$ of vector multiplets. 
For the sake of simplicity, we set Newton's constant  to unity. 
As usual, the K{\" a}hler metric, $G_{a \bar b}$, is defined in terms of the 
K{\" a}hler potential $K(z, \bar z)$ , 
\begin{eqnarray}
e^{-K(z, \bar z)}=-z^{I}N_{IJ}{\overline z}^{J}=\vert X ^{0}\vert ^{-2}, 
\label{eq:kahler}
\end{eqnarray}
where $N_{IJ}$ is related to the second derivative of the prepotential $F$ as
\begin{eqnarray}
N_{IJ}=2\:{\rm Im}F_{IJ}=2\:{\rm Im}\frac{\partial ^{2}F(X)}{
\partial X^{I}\partial X^{J}}.   
\end{eqnarray}
More explicitly, $G_{a\bar b}$ is written
\begin{eqnarray}
G_{a{\bar b}}=\partial _{a}\partial _{\bar b}K
=-\left ( z^{K}N_{KL}\bar z ^{L} \right )^{-1}N_{ab}+
\left ( z^{K}N_{KL}\bar z ^{L}\right )^{-2}N_{aI}\bar z ^{I}N_{bJ}
z^{J}. 
\end{eqnarray}
In (\ref{eq:lagrangian}), we have also introduced the quantity
\begin{eqnarray}
{\cal N}_{IJ}={\overline F}_{IJ}+i\frac{N_{IK}z^{K}N_{JL}z^{L}}{
z^{M}N_{MN}z^{N}}, 
\end{eqnarray}
which is also a function of the moduli. 
The real and imaginary parts of ${\cal N}_{IJ}$ are denoted by 
$\nu _{IJ}$ and $\mu _{IJ}$, respectively; i.e., we have
\begin{eqnarray}
{\cal N}_{IJ}=\nu _{IJ}+i\mu _{IJ}.
\end{eqnarray}

Let us begin with the static metric
\begin{eqnarray}
ds^{2}=-e^{2U(\tau )}(dt)^{2}+e^{-2U(\tau )}
\left \{
\frac{1}{\tau ^{4}}(d\tau )^{2}+
\frac{1}{\tau ^{2}}d\Omega ^{2}
\right \}, 
\label{eq:metric}
\end{eqnarray}
where the horizon corresponds to $\tau \to -\infty$ and the spatial infinity 
to $\tau \to 0$. 
[This metric 
corresponds to the case of extremal black holes, where the non-extremality 
 parameter $c$ has been set equal to zero in (\ref{eq:nonextremalmetric}).]
To solve the equations of motion for $U(\tau )$ and $z^{a}$ coupled to the 
gauge fields, we postulate 
\begin{eqnarray}
F_{tr}^{I}=\tilde q ^{I}, \:\:\:\:\:
F_{\theta \phi }^{I}=p^{I}{\rm sin}\theta , \:\:\:\:\:
G_{I t r}=\tilde p _{I}, \:\:\:\:\:
G_{I \theta \phi }=q_{I}{\rm sin}\theta , 
\label{eq:gaugefields}
\end{eqnarray}
where the magnetic fields are defined as $G_{I \mu \nu}^{-}=\overline{{\cal 
N}}_{IJ}F_{\mu \nu}^{J-}$ and $\tilde q^{I}$ and $\tilde p_{I}$ 
are given by the  electric and magnetic charges $q_{I}$ and $p^{I}$ of  
the black hole as 
\begin{eqnarray}
& & \tilde q^{I}=e^{2U}[(\mu ^{-1})^{IJ}\nu _{JK}p^{K}-(\mu ^{-1})^{IJ}
q_{J}], 
\\
& & \tilde p_{I}=e^{2U}[\nu _{IJ}(\mu ^{-1})^{JK}\nu _{KL}p^{L}-\nu 
_{IJ}(\mu ^{-1})^{JK}q_{K}+\mu _{IJ}p^{J}].
\end{eqnarray}
With the gauge field configurations appearing in
(\ref{eq:gaugefields}), the equations of motion 
turn out to be 
\begin{eqnarray}
& & U''=e^{2U}V_{BH}, 
\label{eq:eqom1}
\\
& & -\left \{ U''-2(U')^{2}
\right \}+2G_{a{\bar b}}(z^{a})'
({\overline z}^{b})'-e^{2U}V_{BH}=0, 
\label{eq:eqom2}
\\
& & \{ G_{a\bar b}({\overline z}^{b})' \}'-\partial _{a}G_{b\bar c}(z^{b})'
({\bar z}^{c})' = e^{2U}\partial _{a}V_{BH}. 
\label{eq:eqom3}
\end{eqnarray}
Here, the black hole potential $V_{BH}$ is given by 
\begin{eqnarray}
V_{BH}(z, \bar z, p, q)=-\frac{1}{2}(p^{I}, q_{J})
\begin{pmatrix}
(\nu \mu ^{-1}\nu +\mu)_{IK} & -{(\nu \mu ^{-1})_{I}}^{L}
\\
-{(\mu ^{-1}\nu )^{J}}_{K} & (\mu ^{-1})^{JL}
\end{pmatrix}
\begin{pmatrix}
p^{K}
\\
q_{L}
\end{pmatrix}
. 
\label{eq:bhpotential}
\end{eqnarray}

Note that the equations of motion (\ref{eq:eqom1})$-$(\ref{eq:eqom3}) can be derived from the Lagrangian 
\begin{eqnarray}
{\cal L}(U(\tau), z(\tau ), \bar z (\tau))=
(U')^{2}+G_{a{\bar b}}(z^{a})'(\overline{z}^{b})'+e^{2U}
V_{BH}(z, \overline{z}, p, q), 
\label{eq:effectivelagrangian}
\end{eqnarray}
supplemented by the constraint 
\begin{eqnarray}
(U')^{2}
+G_{a{\bar b}}(z^{a})'(\overline{z}^{b})'-e^{2U}V_{BH}
(z, \overline{z}, p, q)=0. 
\label{eq:constraint}
\end{eqnarray}
It should also be mentioned that   (\ref{eq:bhpotential}) can be 
expressed as 
\begin{eqnarray}
V_{BH}(z, \bar z, p, q)=\vert Z\vert ^{2} +\vert {\cal D}_{a}Z\vert ^{2}, 
\end{eqnarray}
where 
$Z$, defined by
\begin{eqnarray}
Z=e^{K/2}\left (  p^{I}F_{I}(z)-q_{I}z^{I}  \right ),
\label{eq:central}
\end{eqnarray}
becomes the central charge at the spatial infinity. Note that 
${\cal D}_{a}$ is the K{\" a}hler covariant derivative, i.e., 
${\cal D}_{a}Z=(\partial _{a}+ \frac{\displaystyle{1}}{
\displaystyle{2}}\partial _{a}K )Z$.

\section{The exact solutions }\label{sec:exact}
\setcounter{equation}{0}
\setcounter{figure}{0}
\renewcommand{\theequation}{3.\arabic{equation}}
\renewcommand{\thefigure}{3.\arabic{figure}}

Let us consider the simple case of the STU model, whose prepotential 
is given by 
\begin{eqnarray}
F=-\frac{X^{1}X^{2}X^{3}}{X^{0}}.
\label{eq:fxxx}
\end{eqnarray}
Here, we would like to solve   the equations of motion for the 
D0-D4 charge configuration ($q_{0}, p^{1}, p^{2}, p^{3}$).
We consider the case in which the moduli fields 
are  complex numbers, i.e., 
\begin{eqnarray}
z^{a}=x+iy. \hskip0.5cm (a=1,2,3)
\label{eq:zaarecommon}
\end{eqnarray}
Note, however, that  we are considering the special case in which 
the real and imaginary parts,  $x$ and $y$,  are assumed to 
be common to the three scalar fields.
The charges $p^{a}\:\:(a=1, 2,3)$ are also assumed  to be the same, for simplicity:
\begin{eqnarray}
p^{1}=p^{2}=p^{3}=p.  
\end{eqnarray}
Under these assumptions, the K{\" a}hler potential is computed 
according to (\ref{eq:kahler}), and it is found to be 
\begin{eqnarray}
e^{K}=\frac{1}{8y^{3}}.
\end{eqnarray}
The K{\" a}hler metric $G_{a \bar b}$ and the quantities $\nu _{IJ}$ and $\mu _{IJ}$
are given by
\begin{eqnarray}
G_{a\bar b}=\frac{1}{4y^{2}}
\begin{pmatrix}
1 & 0 & 0
\\
0 & 1 & 0
\\
0 & 0 & 1
\end{pmatrix}, 
\end{eqnarray}
\begin{eqnarray}
\nu _{IJ}=
\begin{pmatrix}
-2x^{3} &  x^{2}  & x^{2}  & x^{2} 
\\
x^{2} & 0 & -x & -x 
\\
x^{2} & -x & 0 & -x
\\
x^{2} & -x & -x & 0
\end{pmatrix}
, 
\end{eqnarray}
and
\begin{eqnarray}
\mu_{IJ}=
y 
\begin{pmatrix}
-3x^{2}-y^{2} &  x  & x  & x 
\\
x & -1 & 0 & 0 
\\
x & 0 & -1 & 0
\\
x & 0 & 0 & -1
\end{pmatrix}
.
\end{eqnarray}
It is also straightforward to obtain the black hole potential 
(\ref{eq:bhpotential}) 
and the Lagrangian (\ref{eq:effectivelagrangian})
in explicit form:

\begin{eqnarray}
V_{BH}&=&\frac{1}{2}\left ( 
3p^{2}y+\frac{q_{0}^{2}}{y^{3}}+9p^{2}\frac{x^{4}}{y^{3}}
+12p^{2}\frac{x^{2}}{y}+6pq_{0}\frac{x^{2}}{y^{3}}
\right ), 
\\
{\cal L}&=&(U')^{2}+\frac{3}{4y^{2}}\left \{
(y')^{2}+(x')^{2}
\right \}+e^{2U}V_{BH}.
\end{eqnarray}
The equations of motion that we have to solve are obtained by varing ${\cal L}$ with respect to $U$, $x$ and $y$. This yields
\begin{eqnarray}
U''
&=&
\frac{1}{2}e^{2U}\left (
3p^{2}y+\frac{q_{0}^{2}}{y^{3}}+9p^{2}\frac{x^{4}}{y^{3}}+
12p^{2}\frac{x^{2}}{y}+6pq_{0}\frac{x^{2}}{y^{3}}
\right ), 
\label{eq:u}
\\
\left (\frac{x'}{y^{2}}\right )'
&=&
4e^{2U}\left (
3p^{2}\frac{x^{3}}{y^{3}}+2p^{2}\frac{x}{y}+pq_{0}\frac{x}{y^{3}}
\right ), 
\label{eq:x}
\\
\frac{y''}{y^{2}}-\frac{(y')^{2}}{y^{3}}+\frac{(x')^{2}}{y^{3}}
&=&
e^{2U}
\left ( p^{2}-\frac{q_{0}^{2}}{y^{4}}-9p^{2}\frac{x^{4}}{y^{4}}
-4p^{2}\frac{x^{2}}{y^{2}}-6pq_{0}\frac{x^{2}}{y^{4}}
\right ).
\label{eq:y}
\end{eqnarray}

Now, in order to rearrange (\ref{eq:u})$-$({\ref{eq:y}), we first absorb 
$p$ and $q_{0}$  
by introducing $M_{0}$ and the new variables $\xi $ and $\phi$ via   
\begin{eqnarray}
M_{0}^{2}=2\sqrt{p^{3}q_{0}}, \:\:\:
x^{2}=\frac{q_{0}}{p} \xi , \;\;\; 
y=\sqrt{\frac{q_{0}}{p}}e^{\phi}. 
\label{eq:xandy}
\end{eqnarray}
Here, we assume $q_{0}/p > 0$ \cite{tt}. 
We then find that the constraint (\ref{eq:constraint}) and 
the equations of motion  
(\ref{eq:u}), (\ref{eq:x}), and ({\ref{eq:y}) are given by
\begin{eqnarray}
& &
(U')^{2}+\frac{3}{4}(\phi ')^{2}+\frac{3(\xi ')^{2}}{16\xi }e^{-2\phi }
=\frac{M_{0}^{2}}{4}e^{2U}\left \{
3e^{\phi}(1+4\xi e^{-2\phi})+e^{-3\phi }(1+3\xi )^{2}
\right \}, 
\nonumber\\
\label{eq:constraint2}
\\
& &
U''=\frac{M_{0}^{2}}{4}e^{2U}\left \{
3e^{\phi}(1+4\xi e^{-2\phi})+e^{-3\phi }(1+3\xi )^{2}
\right \}, 
\label{eq:u2}
\\
& &
\frac{1}{2\xi }\left ( \xi ' e^{-2\phi} \right )'-\frac{(\xi ')^{2}}{
4 \xi ^{2}}e^{-2\phi }=M_{0}^{2}e^{2U}\left \{
4e^{-\phi}+2e^{-3\phi }(1+3\xi)
\right \}, 
\label{eq:x2}
\\
& & 
\phi ''+\frac{(\xi ')^{2}}{4\xi }e^{-2\phi}=\frac{M_{0}^{2}}{2}
e^{2U}\left \{ 
e^{\phi}-4\xi e^{-\phi}-e^{-3\phi }(1+3\xi )^{2}
\right \}.
\label{eq:y2}
\end{eqnarray}

Next, we consider the linear combinations of $U$ and $\phi $ 
\begin{eqnarray}
\alpha =U+\frac{1}{2}\phi, \hskip0.5cm \beta =U-\frac{3}{2}\phi ,
\label{eq:linearcombination}
\end{eqnarray}
with which    (\ref{eq:u2}) and (\ref{eq:y2}) are written
\begin{eqnarray}
\alpha '' + \frac{(\xi ')^{2}}{8\xi }e^{\beta -\alpha}&=&M_{0}^{2}e^{2\alpha }
\left (
2\xi e^{\beta -\alpha }+1
\right ), 
\label{eq:alpha}
\\
\beta ''-\frac{3}{4}\left ( \xi ' e^{\beta -\alpha }\right )'
&=& M_{0}^{2}e^{2\beta }(3\xi +1), 
\label{eq:beta}
\end{eqnarray}
with the help  of (\ref{eq:x2}). 
Also, to eliminate $\xi '$, we combine (\ref{eq:constraint2}) and 
(\ref{eq:alpha}):
\begin{eqnarray}
3(\alpha ')^{2}+(\beta ')^{2}-6\alpha ''=M_{0}^{2}\left \{
-3e^{2\alpha }+e^{2\beta }(1+3\xi )^{2}
\right \}.
\label{eq:constraint3}
\end{eqnarray}

We now postulate that the solution $\alpha (\tau)$ 
coincides with what we would have obtained in the case $\xi=0$. In other words, 
we assume  
\begin{eqnarray}
e^{-\alpha }=\alpha _{0}-M_{0}\tau ,
\label{eq:alphaansatz}
\end{eqnarray}
where $\alpha _{0}$ is an arbitrary constant. 
This postulate enables us to write (\ref{eq:alpha}) and 
(\ref{eq:constraint3}) as the first-order differential equations
\begin{eqnarray}
\xi '=4M_{0}e^{\alpha }\xi =\frac{4M_{0}\xi }{\alpha _{0}-M_{0}\tau }, 
\hskip1cm
\beta '=M_{0}e^{\beta }(3\xi  +1).
\label{eq:xibeta}
\end{eqnarray}
In fact, we can confirm that $\xi$ and $\beta$ given by (\ref{eq:xibeta}) also satisfy 
the second-order equation 
(\ref{eq:beta}). 
It is now almost trivial to work out the solution
\begin{eqnarray}
\xi &=&
\frac{\gamma _{0}^{2}}{(\alpha _{0}-M_{0}\tau )^{4}}, 
\label{eq:558}
\\
e^{-\beta }&=&
(\beta _{0}-M_{0}\tau )-\frac{\gamma _{0}^{2}}{(\alpha _{0}-M_{0}\tau )^{3}}. 
\label{eq:562}
\end{eqnarray}
Here, $\beta _{0}$ and $\gamma _{0}$ are arbitrary constants. 
Combining (\ref{eq:alphaansatz}) and (\ref{eq:562}),
we arrive at
\begin{eqnarray}
& & e^{2\phi}=e^{\alpha -\beta}=
\frac{\beta _{0}-M_{0}\tau }{\alpha _{0}-M_{0}\tau }-\frac{
\gamma _{0}^{2}}{(\alpha _{0}-M_{0}\tau )^{4}}, 
\label{eq:exp2phi}
\\
& & e^{-4U}=e^{-3\alpha -\beta}=
(\alpha _{0}-M_{0}\tau )^{3}(\beta _{0}-M_{0}\tau )-\gamma _{0}^{2}. 
\label{eq:exp-4u}
\end{eqnarray}
We can easily confirm that 
(\ref{eq:558}), (\ref{eq:exp2phi}), and (\ref{eq:exp-4u}) are consistent with the original equations (\ref{eq:constraint2})$-$(\ref{eq:y2}).

At the spatial infinity ($\tau \to 0$), we set  $U(0)=0$ and 
$U'(0)=M$, where $M$ is the black hole mass. 
These yield the constraint
\begin{eqnarray}
\alpha _{0}^{3}\beta _{0}-\gamma _{0}^{2}=1 ,  
\end{eqnarray}
together with  the black hole mass 
\begin{eqnarray}
M=\frac{M_{0}}{4}\left (
\alpha _{0}^{3}+3\alpha _{0}^{2}\beta _{0}
\right ). 
\end{eqnarray}
This deviates from $M_{0}$ in a manner that depends on the initial conditions for 
the moduli fields at the spatial infinity. 
It should also be  remarked that 
our solution is non-BPS, because (\ref{eq:central}) in our case 
is
\begin{eqnarray}
Z=\frac{1}{\sqrt{8y^{3}}}\left \{
-3p(x+iy)^{2}-q_{0}
\right \}, 
\end{eqnarray}
and at the spatial infinity we have
\begin{eqnarray}
\lim _{\tau \to 0}\vert Z \vert =\frac{M_{0}}{4}
\sqrt{\left ( \alpha _{0}^{3}+3\alpha _{0}^{2}\beta _{0}
\right )^{2}-12\alpha _{0}^{2}}
\neq M .
\label{eq:nonbpsdesu}
\end{eqnarray}

It is illuminating to  express our solution  in terms of the harmonic functions
\begin{eqnarray}
H=\frac{p}{M_{0}}\left (
\alpha _{0}-M_{0}\tau \right ), 
\hskip0.5cm
\tilde H_{0}=\frac{q_{0}}{M_{0}}\left (
\beta _{0}-M_{0}\tau \right ). 
\label{eq:harmonic}
\end{eqnarray}
In terms of these,
(\ref{eq:558}), (\ref{eq:exp2phi}), and (\ref{eq:exp-4u}) 
can be rewritten  as 
\begin{eqnarray}
z&=&\pm \sqrt{\frac{q_{0}}{p}}\sqrt{\xi} +i\sqrt{\frac{q_{0}}{p}}
e^{\phi }
=\pm \frac{\gamma _{0}}{2H^{2}}+i\sqrt{\frac{\tilde H_{0}}{H}-
\frac{\gamma _{0}^{2}}{4H^{4}}}, 
\label{eq:z}
\\
e^{-2U}&=&\sqrt{4H^{3}\tilde H_{0}-\gamma _{0}^{2}}.
\label{eq:expminusuu}
\end{eqnarray}
Note that the moduli fields are attracted to the purely imaginary number 
\begin{eqnarray}
z \vert _{\rm horizon} =i\sqrt{\frac{q_{0}}{p}} 
\label{eq:moduliathorizon}
\end{eqnarray}
at the horizon ($\tau \to -\infty$). 
Apparently, the attractor mechanism is effective, 
as 
(\ref{eq:moduliathorizon})
is  independent 
of the initial conditions, $\alpha _{0}$, $\beta _{0}$, and $\gamma _{0}$.
We also note that  the black hole potential 
$V_{BH}$ is equal to $M_{0}^{2}$ at the horizon, 
and the black hole entropy 
$S_{\rm BH}$ is given in terms of the 
charges carried by the black hole alone: 
$S_{\rm BH}=\pi V_{BH}\vert _{\rm horizon}=
\pi M_{0}^{2}=2\pi \sqrt{p^{3}q_{0}}$.

Finally, the following 
 fact should be pointed out.
If we replace the charges in (\ref{eq:moduliathorizon})
by the corresponding harmonic functions (\ref{eq:harmonic}), 
then  we  obtain 
$i\sqrt{\tilde H_{0}/H}$.  
The solution (\ref{eq:z}), however, posseses the additional 
 parameter $\gamma _{0}$, 
which comes from the initial condition at the infinity for  the real
 part of the moduli fields.

\section{Summary}\label{}
\setcounter{equation}{0}
\setcounter{figure}{0}
\renewcommand{\theequation}{5.\arabic{equation}}
\renewcommand{\thefigure}{5.\arabic{figure}}

In the present paper we have discussed four-dimensional 
${\cal N}=2$ supergravity from the Type IIA superstring, which is 
described by (\ref{eq:lagrangian}). 
We have solved the equations of motion for the STU model 
(\ref{eq:fxxx}) with D0-D4 brane charges, and our solutions are 
summarized in 
(\ref{eq:z}) and (\ref{eq:expminusuu}).
It should be stressed that our 
moduli fields are complex, in contrast to those of previous works. 
This  shows that even if we  replace the charges in 
(\ref{eq:moduliathorizon}) by the corresponding harmonic functions, 
we would not get back the general solutions (\ref{eq:z}) 
for $\gamma _{0}\neq 0$.

As we see in (\ref{eq:moduliathorizon}), 
the moduli fields are attracted at the horizon to a value determined 
by only the charges of the black hole. In other words, they are independent 
of the values of $\alpha _{0}$, $\beta _{0}$, and $\gamma _{0}$. Note that 
our solutions are for the non-BPS black hole,
as mentioned in (\ref{eq:nonbpsdesu}).


Throughout the present paper, we have been mostly concerned with 
the analytic behavior of the moduli fields 
in the whole $\tau$-space. It is known 
that for BPS black holes we can avoid the problem of solving the differential 
equations themselves to determine the $\tau$-space behavior. In that case, it is necessary only to solve the algebraic ``generalized 
stabilization  equation," in which all of the non-vanishing charges 
$(q_{0}, q_{a}, p^{0}, p^{a})$ in the attractor equation are replaced 
by the corresponding harmonic functions (see Appendix \ref{bps}). 
For non-BPS black holes, however, it is yet to be confirmed 
that such a simple algebraic prescription 
is equivalent to solving the second-order differential equations. 
We hope our solutions for the D0-D4 brane system 
with complex moduli fields is useful for gaining insight into 
this and related problems.

After completing the present work, the authors were informed of the 
recent interesting paper by Cardoso et al \cite{ccdo}. They made use of the 
first-order flow equation \cite{cersole} 
and have obtained exact solutions 
for complex scalar fields in the whole space.

\section*{Acknowledgements}

We would like to thank Dr.  Yoshifumi Hyakutake for useful discussions. 
We are also grateful to Professor Gabriel Cardoso for calling our attention to 
\cite{ccdo}.  
The  work of T.K. is partially supported  
by a Grant-in-Aid from the Ministry of Education (No. 19034006). 
K.H. is supported in part by JSPS Research Fellowship 
for Young Scientists.

\appendix

\section{The Generalized Stabilization Equation
and  the BPS full solution}\label{bps}
\setcounter{equation}{0}
\setcounter{figure}{0}
\renewcommand{\theequation}{A.\arabic{equation}}
\renewcommand{\thefigure}{A.\arabic{figure}}


We would like to supplement the analysis of extremal BPS black holes given in the main text with a treatment of the BPS attractor equations. The purpose of this calculation 
is to show that if we replace the charges in the double extremal solutions
by the corresponding harmonic functions, then we obtain solutions to the 
first-order BPS condition. As a byproduct of this analysis, 
we are able to incorporate 
 the real part of the moduli fields.

The BPS black hole solution with non-vanishing charges 
$(p^{a}, p^{0}, q_{a}, q_{0})$ $(a=1,2, \cdots , N_{V})$ 
is derived from the generalized stabilization equation 
 
\begin{eqnarray}
i
\begin{pmatrix}
H^{0}
\\
H^{a}
\\
\tilde H_{0}
\\
\tilde H_{a}
\end{pmatrix}
=e^{K/2}\left \{
\bar \Sigma 
\begin{pmatrix}
1
\\
z^{a}
\\
F_{0}
\\
F_{a}
\end{pmatrix}
-\Sigma 
\begin{pmatrix}
1
\\
\bar z^{a}
\\
\bar F_{0}
\\
\bar F_{a}
\end{pmatrix}
\right \},
\label{eq:appbpsattractor}
\end{eqnarray}
which has been obtained by replacing the charges in the 
BPS  attractor equation by the corresponding harmonic functions
\cite{sabra}, \cite{denef}, \cite{mohaupt}.
Here, the harmonic functions are defined by 
\begin{eqnarray}
H^{a}=h^{a}-p^{a}\tau, \hskip0.3cm  
H^{0}=h^{0}-p^{0}\tau , \hskip0.3cm
\tilde H_{a}=\tilde h_{a}-q_{a}\tau, \hskip0.3cm  
\tilde H_{0}=\tilde h_{0}-q_{0}\tau .
\end{eqnarray}
We have also substituted the harmonic functions for the charges in 
(\ref{eq:central}), and thereby defined the following:  
\begin{eqnarray}
\Sigma \equiv e^{K/2}\left (
H^{I}F_{I}-\tilde H_{I}z^{I}
\right ).
\end{eqnarray}
The solution of the metric is given by (\ref{eq:metric}) with
\begin{eqnarray}
e^{-2U}=\vert \Sigma \vert ^{2}. 
\label{eq:appmetric}
\end{eqnarray}
In addition, we have to impose the condition of the asymptotic flatness of the metric 
at the spatial infinity,
\begin{eqnarray}
e^{-2U(\tau=0)}=1, 
\end{eqnarray}
and the constraint 
\begin{eqnarray}
h^{I}q_{I}=\tilde h_{I}p^{I}. 
\end{eqnarray}

We now confine ourselves to the $N_{V}=3$ case 
and adopt the prepotential (\ref{eq:fxxx}). 
We further  assume that 
$p^{a}$, $ q_{a}$, $h^{a}$, and $\tilde h_{a}$ are 
 common to  the three scalar fields,  i.e., 
\begin{eqnarray}
H^{1}=H^{2}=H^{3}=H\equiv h-p\tau , 
\\
\tilde H_{1}=\tilde H_{2} =\tilde H_{3}
=\tilde H \equiv \tilde h - q\tau .
\end{eqnarray}
We also assume that the three scalar functions take the same forms 
as in (\ref{eq:zaarecommon}).
It is then straightforward to solve the algebraic equation  
(\ref{eq:appbpsattractor}), and we get
\begin{eqnarray}
x&=&-\frac{1}{2}\frac{H\tilde H+H^{0}\tilde H_{0}}{H^{2}+\tilde HH^{0}}, 
\label{eq:xsol}
\\
y&=&\sqrt{\frac{\tilde H^{2}-H\tilde H_{0}}{H^{2}+\tilde HH^{0}}-
\frac{1}{4}\left (
\frac{H\tilde H+H^{0}\tilde H_{0}}{H^{2}+\tilde HH^{0}}
\right )^{2}}  . 
\label{eq:ysol}
\end{eqnarray}
According to (\ref{eq:appmetric}), the metric is given in terms of 
\begin{eqnarray}
e^{-2U}=2y(H^{2}+\tilde HH^{0}). 
\label{eq:usol}
\end{eqnarray}
We have confirmed by explicit calculation that the solutions 
(\ref{eq:xsol}), (\ref{eq:ysol}), and (\ref{eq:usol}) in fact satisfy the 
first-order BPS condition  
\cite{springer}, 
\begin{eqnarray}
U'=e^{U}\vert Z \vert , \hskip1cm (z^{a})'=e^{U}G^{a \bar b}
{\cal D}_{\bar b}\bar Z
\frac{Z}{\vert Z \vert }. 
\end{eqnarray}

A remark is in order with regard to the D0-D4 system, i.e., 
$p^{0}=0$,   $q_{a}=0$ and $q_{0}/p < 0$. In this case,  
 (\ref{eq:xsol}) and (\ref{eq:ysol}) are
\begin{eqnarray}
x=-\frac{1}{2}\frac{H\tilde h +h^{0}\tilde H_{0}}{H^{2}+\tilde h h^{0}}, 
\hskip0.5cm
y=\sqrt{\frac{\tilde h^{2}-H\tilde H_{0}}{H^{2}+\tilde h h^{0}}
-\frac{1}{4}\left (
\frac{H\tilde h+h^{0}\tilde H_{0}}{H^{2}+\tilde h h^{0}}
\right )^{2}
},
\label{eq:hotta}
\end{eqnarray}
and at the horizon, these become
\begin{eqnarray}
\lim _{\tau \to -\infty}x=0, 
\hskip0.5cm
\lim _{\tau \to -\infty }y=\sqrt{-\frac{q_{0}}{p}.
}
\label{eq:app847}
\end{eqnarray}
It is worth mentioning that the solutions 
(\ref{eq:hotta})
  have more parameters than
what we would obtain  by replacing the charges in (\ref{eq:app847}) 
by the corresponding harmonic functions.
This is similar to the situation for non-BPS black holes 
discussed in \S \ref{sec:exact}.

\section{The Non-Extremal Black Holes}\label{}
\setcounter{equation}{0}
\setcounter{figure}{0}
\renewcommand{\theequation}{B.\arabic{equation}}
\renewcommand{\thefigure}{B.\arabic{figure}}

A close inspection of the black hole potential $V_{BH}$ reveals 
that the attractor mechanism is ineffective in some cases, in particular 
for non-extremal black holes \cite{gijt}, \cite{tt}, \cite{nonbpsattractor}.
This fact is confirmed in \cite{gijt}, in which  
explicit solutions in a particular model are employed. 
For the sake of completeness, 
here we present similar analysis adapted to the general prepotential 
\begin{eqnarray}
F(X)=D_{abc}\frac{X^{a}X^{b}X^{c}}{X^{0}}.
\hskip1cm (a,b,c=1, \cdots , N_{V})
\label{eq:generalf}
\end{eqnarray}
The fact that the attractor mechanism is ineffective is shown here for non-extremal black holes 
with D0-D4 brane charges on the basis of the exact solutions.

We begin with the metric ansatz for the non-extremal case,
\begin{eqnarray}
ds^{2}=-e^{2U(\tau )}(dt)^{2}+e^{-2U(\tau )}
\left \{
\frac{c^{4}}{{\rm sinh}^{4}c\tau }(d\tau )^{2}+
\frac{c^{2}}{{\rm sinh}^{2}c\tau }d\Omega ^{2}
\right \}, 
\label{eq:nonextremalmetric}
\end{eqnarray}
where a non-vanishing value of $c$ implies that this is the non-extremal black hole.
We use the general prepotential (\ref{eq:generalf})
and  consider the D0-D4 brane charges 
$(q_{0}, p^{a})$, while $q_{a}=p^{0}=0$.
We also   confine ourselves to   purely imaginary moduli fields and set
\begin{eqnarray}
z^{a}=ip^{a}\sqrt{\frac{q_{0}}{D}}e^{\phi}, 
\hskip0.5cm 
({\rm for}\:\:\:q_{0} < 0)
\label{eq:nonextremalbps}
\\
z^{a}=ip^{a}\sqrt{\frac{-q_{0}}{D}}e^{\phi}, 
\hskip0.5cm 
({\rm for}\:\:\:q_{0} > 0)
\label{eq:nonextremalnonbps}
\end{eqnarray}
where $D=D_{abc}p^{a}p^{b}p^{c}<0$. 
Note that in the extremal limit ($c\to 0 $),  
  (\ref{eq:nonextremalbps})
corresponds to the BPS case and  (\ref{eq:nonextremalnonbps}) to 
the non-BPS case \cite{tt}.

The equations of motion are  
\begin{eqnarray}
& & (U')^{2}+\frac{3}{4}(\phi ')^{2}=\frac{M_{0}^{2}}{4}
e^{2U}\left (3e^{\phi}+e^{-3 \phi}\right )+c^{2},
\label{eq:nonc}
\\
\label{eq:nonextu}
& & U''=\frac{M_{0}^{2}}{4}e^{2U}\left ( 3e^{\phi}+e^{-3\phi} \right ), 
\\
&&
\phi ''=\frac{M_{0}^{2}}{2}e^{2U}\left(   e^{\phi}-e^{-3\phi} \right )
\end{eqnarray}
for both (\ref{eq:nonextremalbps}) and  (\ref{eq:nonextremalnonbps}) . 
Here, we have defined 
\begin{eqnarray}
M_{0}^{2}=2\sqrt{q_{0}D}, \hskip0.5cm ({\rm for}\:\:\: q_{0} < 0)
\\
M_{0}^{2}=2\sqrt{-q_{0}D}. \hskip0.5cm ({\rm for}\:\:\: q_{0} > 0)
\end{eqnarray}
As we see, the effects of the non-extremal nature of the black hole appear only in (\ref{eq:nonc}).

We next define new variables $\alpha $
and $\beta$ as in (\ref{eq:linearcombination}). Their equations of motion are easily found to be
\begin{eqnarray}
\alpha ''=M_{0}^{2}e^{2\alpha }, \hskip0.5cm 
\beta ''=M_{0}^{2}e^{2\beta }. 
\label{eq:appalphabeta}
\end{eqnarray}
The constraint equation is also expressed as 
\begin{eqnarray}
\frac{3}{4}(\alpha ')^{2}+\frac{1}{4}(\beta ')^{2}=\frac{M_{0}^{2}}{
4}\left ( 3e^{2\alpha }+e^{2\beta } \right )+c^{2}. 
\label{eq:appconstraint}
\end{eqnarray}
The equations in (\ref{eq:appalphabeta})  are of the Toda-type, and their 
solutions are
\begin{eqnarray}
e^{-\alpha }=\frac{{\rm sinh} [A(\alpha _{0}-M_{0}\tau)]}{A}, 
\hskip0.5cm
e^{-\beta }=\frac{{\rm sinh} [B(\beta _{0}-M_{0}\tau)]}{B}. 
\end{eqnarray}
Here, $A$, $B$, $\alpha _{0}$, and $\beta _{0}$ are constants of integration. 
These solutions, in turn, give $U$ and $\phi $ through the following:
\begin{eqnarray}
e^{-U}&=&
\left (
\frac{{\rm sinh}[A(\alpha _{0}-M_{0}\tau )]}{A}
\right )^{3/4}
\left (
\frac{{\rm sinh}[B(\beta _{0}-M_{0}\tau )]}{B}
\right )^{1/4}, 
\label{eq:appexpu}
\\
e^{\phi }&=&
\left (
\frac{A}{B}\frac{{\rm sinh}[B(\beta _{0}-M_{0}\tau )]}
{{\rm sinh}[A(\alpha  _{0}-M_{0}\tau )]}
\right )^{1/2}.
\label{eq:appexpphi}
\end{eqnarray}

The constraint equation (\ref{eq:appconstraint}) imposes a relation between 
the integration constants $A$ and $B$:
\begin{eqnarray}
\frac{M_{0}^{2}}{4}\left (  3A^{2}+B^{2} \right )=c^{2}. 
\label{eq:app98}
\end{eqnarray}
We thus see that in the extremal case ($c \to 0$), which implies the limits
$A \to 0 $ and $B \to 0$, the solutions reduce to 
\begin{eqnarray}
e^{-U} & \to & (\alpha _{0}-M_{0}\tau )^{3/4}(\beta _{0}-M_{0}\tau )^{1/4}, 
\\
e^{\phi} & \to &  \sqrt{\frac{\beta _{0}-M_{0}\tau }
{\alpha _{0}-M_{0}\tau }}. 
\end{eqnarray}

The mass of the black hole $M$ is defined by the asymptotic form of 
the metric as $\tau \to 0$:
\begin{eqnarray}
e^{-U(\tau )} \to 1-M\tau .
\end{eqnarray}
This provides the constraint
\begin{eqnarray}
\left ( \frac{{\rm sinh}A\alpha _{0}}{A}\right ) ^{3}
\left ( \frac{{\rm sinh}B\beta _{0} }{B} \right )=1,
\end{eqnarray}
together with the formula for the mass
\begin{eqnarray}
M=\frac{M_{0}}{4}
\left \{
\left (
\frac{{\rm sinh}A\alpha _{0}}{A}
\right )^{3}{\rm cosh}B\beta _{0}+3\left (
\frac{{\rm sinh}A\alpha _{0}}{A}
\right )^{2}\frac{{\rm sinh}B\beta _{0}}{B}
{\rm cosh}A\alpha _{0}
\right \}. 
\nonumber\\
\end{eqnarray}
The minimum value of $M$ is 
\begin{eqnarray}
\frac{M_{0}}{4}\left (
\alpha _{0}^{3}+3\alpha _{0}^{2}\beta _{0}
\right ),
\end{eqnarray}
and this corresponds to the extremal case, i.e.,  $A=B=0$. 

The central charge can also be computed as 
\begin{eqnarray}
Z=\frac{1}{4}M_{0}\left (3e^{\phi /2}\pm e^{-3\phi /2}\right ), 
\end{eqnarray}
where $+$ and $-$ correspond to $q_{0}<0$ and $q_{0}>0$, respectively. 
We can easily confirm that $M\neq \vert Z \vert _{\tau =0}$; i.e.,
our non-extremal solution is  non-BPS.

The behavior of the moduli fields at the horizon ($\tau \to -\infty$)
is seen from (\ref{eq:appexpphi}):
\begin{eqnarray}
e^{\phi }\to \sqrt{\frac{A}{B}}e^{(B\beta _{0}-A\alpha _{0})/2}
e^{(A-B)M_{0}\tau /2}.
\end{eqnarray}
This is divergent if $A<B$ and  vanishing if $A>B$. 
If we impose the condition that the moduli fields are regular and 
non-vanishing, then we have to additionally impose the condition
$A=B$. Apparently, even for $A=B$, the value of the moduli fields at the 
horizon depends on the aribitrary constants $\alpha _{0}$ and $\beta _{0}$. 
Thus, the attractor mechanism is ineffectual.


Finally, let us evaluate the entropy of the black hole. 
For $A=B$,  the limiting behavior
\begin{eqnarray}
e^{-2U}\frac{c^{2}}{{\rm sinh}^{2}c\tau }
&\to &
\left ( \frac{e^{A(\alpha _{0}-M_{0}\tau )}}{2A}\right )^{3/2}
\left ( \frac{e^{A(\beta  _{0}-M_{0}\tau )}}{2A}\right )^{1/2}
\frac{4c^{2}}{e^{-2c\tau}}
\\
&=& M_{0}^{2}e^{A(3\alpha _{0}+\beta _{0})/2} 
\end{eqnarray}
can be derived for $\tau \to -\infty$. 
Here, use has been made of the relation 
$M_{0}^{2}A^{2}=c^{2}$, which comes from (\ref{eq:app98}) by setting $A=B$. 
Thus, the Bekenstein-Hawking entropy, 
\begin{eqnarray}
\frac{1}{4}\times \left ( {\rm Area} \right )=\pi M_{0}^{2}
e^{A(3\alpha _{0}+\beta _{0})/2},
\end{eqnarray}
depends on the constants $\alpha _{0}$ and $\beta _{0}$   
if $A\neq 0$.


\end{document}